\documentclass[preprint,aps,pre]{revtex4}
\usepackage[dvips]{graphicx}
\usepackage{latexsym}

\begin{document}

\title{Static and Dynamic Properties of Stretched Water}

\author{Paulo A. Netz$^1$, Francis W. Starr$^2$,  H. Eugene
Stanley$^3$, and  Marcia C. Barbosa$^{3,4}$}

\affiliation{$^1$ Departamento de Qu\'{\i}mica, Universidade Luterana do
Brasil,  92420-280, Canoas , RS, Brazil \\
$^2$ Polymers Division and Center for Theoretical and
Computational Materials Science, National Institute of Standards and
Technology, Gaithersburg, MD 20899, USA \\
$^3$ Center for Polymer Studies and Department of Physics,
Boston University, Boston, MA 02215, USA \\
$^4$ Instituto de F\'{\i}sica, Universidade Federal do Rio
Grande do Sul, Caixa Postal 15051, 91501-970, Porto Alegre,
RS, Brazil }


\begin{abstract}

We present the results of molecular dynamics simulations of the extended
simple point charge (SPC/E) model of water to investigate the
thermodynamic and dynamic properties of stretched and supercooled water. 
We  locate the
liquid-gas spinodal, and confirm  that the spinodal pressure increases
monotonically with $T$, supporting thermodynamic scenarios for the phase
behavior of supercooled water involving a ``non-reentrant'' spinodal.
The dynamics at negative pressure show a minimum in the diffusion
constant $D$ when the density is decreased at constant temperature,
complementary to the known maximum of $D$ at higher pressures.  We
locate the loci of minima of $D$ relative to the spinodal, showing that
the locus is  inside the thermodynamicaly metastable regions of the
phase-diagram.  These dynamical results reflect the initial enhancement
and subsequent breakdown of the tetrahedral structure and of the
hydrogen bond network as the density decreases. 
\end{abstract}

\maketitle

\section{Introduction}

Water is an important liquid in nature, and is also fundamental in
chemical and technological applications. Although the individual water
molecule has a simple chemical structure, water is considered a complex
fluid because of its anomalous behavior
\cite{debe96,ange82,four97,sta97a,lan82}.  It expands on freezing and,
at a pressure of 1 atm, the density has a maximum at
4$^\circ$C. Additionally, there is a minimum of the isothermal
compressibility at 46$^\circ$C and a minimum of the isobaric heat
capacity at 35$^\circ$C \cite{dou98}. These anomalies are linked with
the microscopic structure of liquid water, which can be regarded as a
transient gel---a highly associated liquid with strongly directional
hydrogen bonds \cite{geig79,Stanley80}. Each water molecule acts as both
a donor and an acceptor of bonds, generating a structure that is locally
ordered, similar to that of ice, but maintaining the long-range disorder
typical of liquids. Despite the extensive work that has been done on
water, many aspects of its behavior remain unexplained.

Several scenarios have been proposed to account for the the anomalous
behavior of the thermodynamic response functions on cooling, each
predicting a different behavior for the liquid spinodal, the line of the
limit of stability separating the region where liquid water is
metastable from the region where the liquid is unstable. (i) According
the stability-limit conjecture \cite{spe82b,spe87}, the pressure of the
spinodal line should decrease on cooling, become negative, and increase
again after passing through a minimum. It reenters the positive pressure
region of the phase diagram at a very low temperature, thereby giving
rise to a line of singularities in the positive pressure region, and
consequently the increase in the thermodynamic response functions on
cooling in the anomalous region is due to the proximity of this
reentrant spinodal. (ii) The critical point hypothesis
\cite{pses,mishima94,critical-point,ms98,tana96,t35}, proposes a new
critical point at the terminus of a first-order phase transition line
separating two liquid phases of different density. The anomalous
increases of the response functions, compressibility, specific heat, and
volume expansivity, is interpreted in terms of this critical
point. (iii) The singularity-free hypothesis
\cite{Stanley80,sas96,sas98} proposes that actually there is no
divergence close to the anomalous region; the response functions grow by
lowering temperature but remain finite, attaining maximum values. Both
hypotheses are consistent with a non-reentrant spinodal, and so
unambiguously identifying the spinodal can rule out at least on
scenario.

Water properties and anomalies can be strongly influenced by the
physical or chemical properties of the medium
\cite{debe96,ange82,four97,sta97a,net98a,koga98}. The effect not only of
applied pressure, but also of negative pressure (``stretching'') is
remarkable. The study of the behavior of this fluid under negative
pressures is relevant not only from the academic point of view, but also
for realistic systems. For example, negative pressures are observed
\cite{poc95}, and seem to play an important role in the mechanism of
water transport in plants. Therefore, properties that modify the
structure of water, especially if this modification is similar to the
effect of stretching (as is the case in some hydrogels
\cite{net98a}), also influence its dynamical behavior.

Dynamic properties, such as the diffusion constant, have been studied in
detail for water systems at atmospheric and at high positive pressures,
both experimentally \cite{jon76,pri87} as well as by computer
simulations \cite{ram87,sci91,bae94,har97,francesco,sta99e,scala}. The
increase of pressure increases the presence of defects and of
interstitial water molecules in the network \cite{sci91}. They disrupt
the tetrahedral local structure, weakening the hydrogen bonds, and thus
increasing the diffusion constant \cite{sta99e,scala}. However, a
further increase in the pressure leads to steric effects which works in
the direction of lowering the mobility. The interplay of these factors
leads to a maximum in the diffusion constant \cite{sta99e,scala} at some
high density $\rho_{\mbox{\scriptsize max}}$. Above this density (or
corresponding pressure), the diffusion of water is in some sense like
that of a normal liquid, controlled by hindrance, with the hydrogen
bonds playing a secondary role.  However, the behavior at very low
$\rho$ is less well understood.

In this paper, we study how the thermodynamics and the dynamics of
low-temperature water are affected by the decrease of the density. We
perform molecular dynamics (MD) simulations of the SPC/E model of water
in the range 210~K $< T <$ 280~K and 0.825~g/cm$^3 < \rho <$
0.95~g/cm$^3$.  State points in this range have negative pressure, and
are either a metastable stretched liquid, or a phase separated
liquid-gas mixture. Additional details of the simulation procedure can
be found in ref.~\cite{sta99e}.    
 Table~\ref{jobs} shows the
thermodynamic and dynamic properties of the simulated state points as
well as the time needed for equilibration and production runs.  At the
lowest $T$ studied, the production runs are not long enough to reliably
estimate $D$.

\section{Location of the Spinodal}

The effect of negative pressure is still not fully understood, and only
a few experimental works \cite{hen80,GDWA90} and simulations
\cite{pses,tana96,francesco,vai93,sta99e,ge94,err00} have been performed on
transport properties of stretched water.  In this negative pressure
region of the phase diagram the system is metastable, and becomes
unstable beyond the spinodal line, so locating the spinodal we can
ensure that our simulated state points lie in the metastable and not in
the unstable region. Moreover, the shape of the spinodal can test the
stability-limit conjecture against the critical point hypothesis and the
singularity-free interpretation, so we first locate the density and
pressure of the spinodal, which we denote $\rho_{\mbox{\scriptsize
sp}}(T)$ and $P_{\mbox{\scriptsize sp}}(T)$, respectively.

We consider the the $P$-$\rho$ behavior along isotherms.  For each $T$,
there is a minimum in $P$ at $\rho \approx 0.86$~g/cm$^3$.  At lower
$\rho$, $P$ increases, signaling cavitation of the liquid. For densities
below $\rho_{\mbox{\scriptsize sp}}$,
$K_T\equiv\rho^{-1}(\partial\rho/\partial P)_T$ becomes negative, so the
liquid is unstable with respect to the vapor phase.  Therefore, the
locus of infinite compressibility locates the spinodal line, and sets
the density below which our simulated state points are no longer
thermodynamically stable.

\begin{figure}
\includegraphics[width=8.6cm,angle=-90]{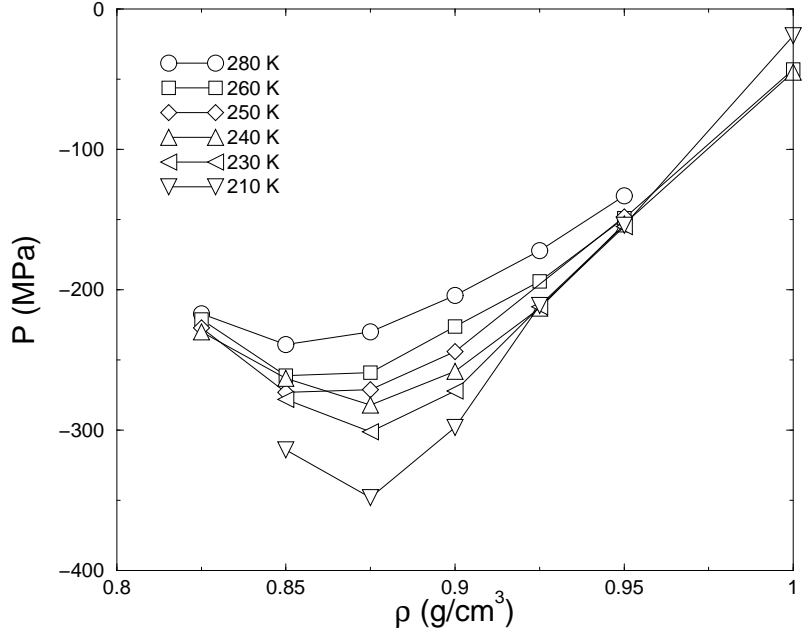}
\caption{Dependence of the pressure on the density, along isotherms, for
the simulations reported here.}
\label{f1}
\end{figure}

\begin{figure}
\includegraphics[width=8.6cm,angle=-90]{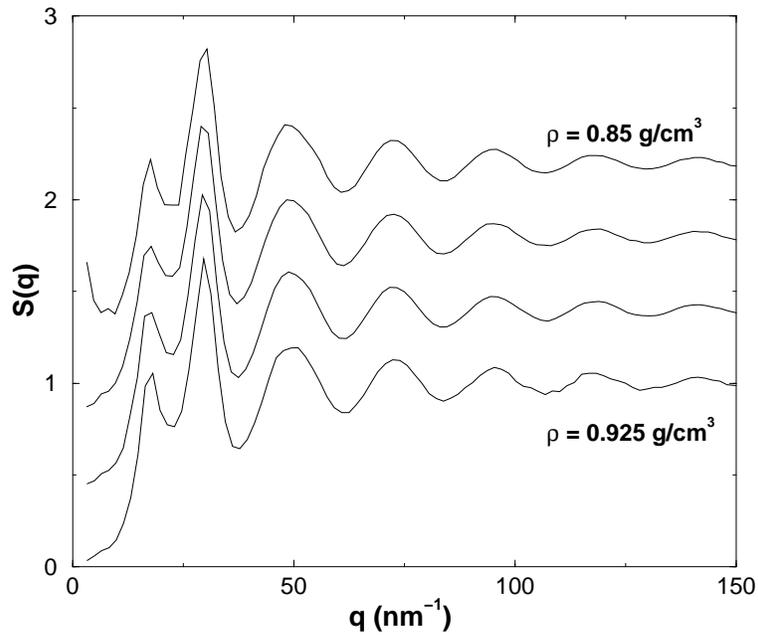}
\caption{At $T=210$~K, the structure factor $S(q)$.  Note the increase 
  of $S(q)$ at small $q$ at $\rho = 0.85$~g/cm$^3$, indicating phase
  separation.}
\label{struct-fact}
\end{figure}

We confirm which state points are ``single phase'' and which are ``phase
separated'' by calculating the oxygen-oxygen partial structure factor
\begin{equation}
S(q) = \frac{1}{N} \sum_{j,k}^N e^{-i{\bf q}\cdot({\bf r}_j-{\bf r}_k)}
\label{eq:s(q)}
\end{equation}
Figure~\ref{struct-fact} shows $S(q)$ at the the lowest $T$ ($T=210$~K)
for each density simulated.  At the lowest density, $\rho =
0.85$~g/cm$^3$, $S(q)$ for $q \lesssim 10$~nm$^{-1}$ increases
significantly, signaling the presence of large-scale structure in the
system due to phase separation of liquid and gas.  Similar behavior is
observed for other state points that would appear to be in the unstable
regime from Fig.~\ref{f1}~\cite{g(r)note}.

\begin{figure}
\includegraphics[width=8.6cm,angle=-90]{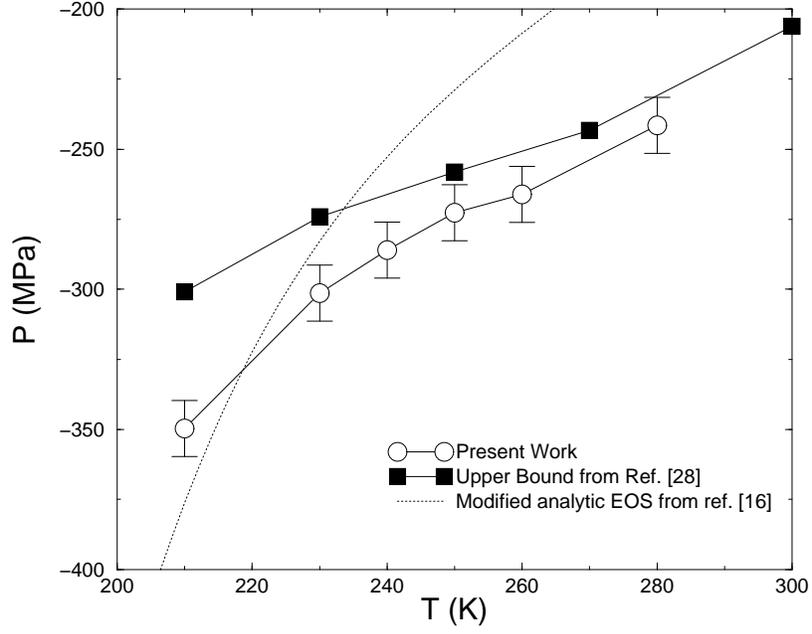}
\caption{Location of the spinodal ($\circ$), showing no evidence for
  re-entrant behavior.  For comparison, we also show the upper bound on
  the spinodal (filled $\Box$) from ref.~\cite{har97} and the estimated
  spinodal (dotted line) from the analytic free energy expression of
  ref.~\cite{t35}. }
\label{f2}
\end{figure}

To test whether the minima in the $P$-$\rho$ isotherms arise from
cavitation of the liquid, we fit at each $T$ a 5th-order polynomial to
the $P$-$\rho$ isotherms, and estimate $P_{\mbox{\scriptsize sp}}(T)$
and $\rho_{\rm sp}(T)$ by the minimum of the polynomial fit.  We find
that $\rho_{\rm sp}(T)$ occurs in the narrow range $0.853 < \rho_{\rm
sp} <0.874$, and so the spinodal density is nearly $T$ independent.  Our
results indicate that $P_{\rm sp}(T)$ decreases monotonically with
decreasing $T$ (Fig.~\ref{f2}), ruling out the possibility of a
re-entrant spinodal in the temperature range studied.

We also compare our estimate of the spinodal with estimates from two
other sources for the SPC/E potential: (i) an estimate of the upper
bound on $P_{\mbox{\scriptsize sp}}(T)$~\cite{har97}, and (ii) the
spinodal estimated by an approximate analytic equation of state
(EOS) \cite{t35}. Figure~\ref{f2} shows that our estimate of
$P_{\mbox{\scriptsize sp}}(T)$ is less than (and hence consistent with)
the upper bound estimate of Ref.~\cite{har97}.  The spinodal obtained
from the EOS in Ref.~\cite{t35} is quantitatively reasonable for $T
\gtrsim 280$~K, but decreases far more quickly with decreasing $T$
(reaching $\approx -2$~GPa at $T=210$~K) than our simulations indicate.
This dramatic decrease likely results from the fact that the fitting
procedure requires $P$ along a reference isotherm, which was performed
at $T=300$~K in the density range $0.95 < \rho < 1.4$~g/cm$^3$.  This
fit is not reliable outside of this density range, where the spinodal
lies, causing an erroneous spinodal estimate.  These errors are
compounded as the fit is pushed further from the reference $T$.  To
improve the estimate, we repeat the fitting procedure using a new
reference isotherm $T=210$~K, and expand the density range down to $\rho
= 0.85$~g/cm$^3$.  The results of this fit are shown in Fig.~\ref{f2},
and we obtain more reasonable agreement with our direct simulation
results.  Ref.~\cite{t35} also used this EOS to predict a liquid-liquid
phase transition for $T\lesssim 130$~K; this prediction should not be
affected by using data at lower $\rho$, since the transition lies within
the density range of the original fits.

\section{Dynamic Properties}

\begin{figure}
\includegraphics[width=8.6cm,angle=-90]{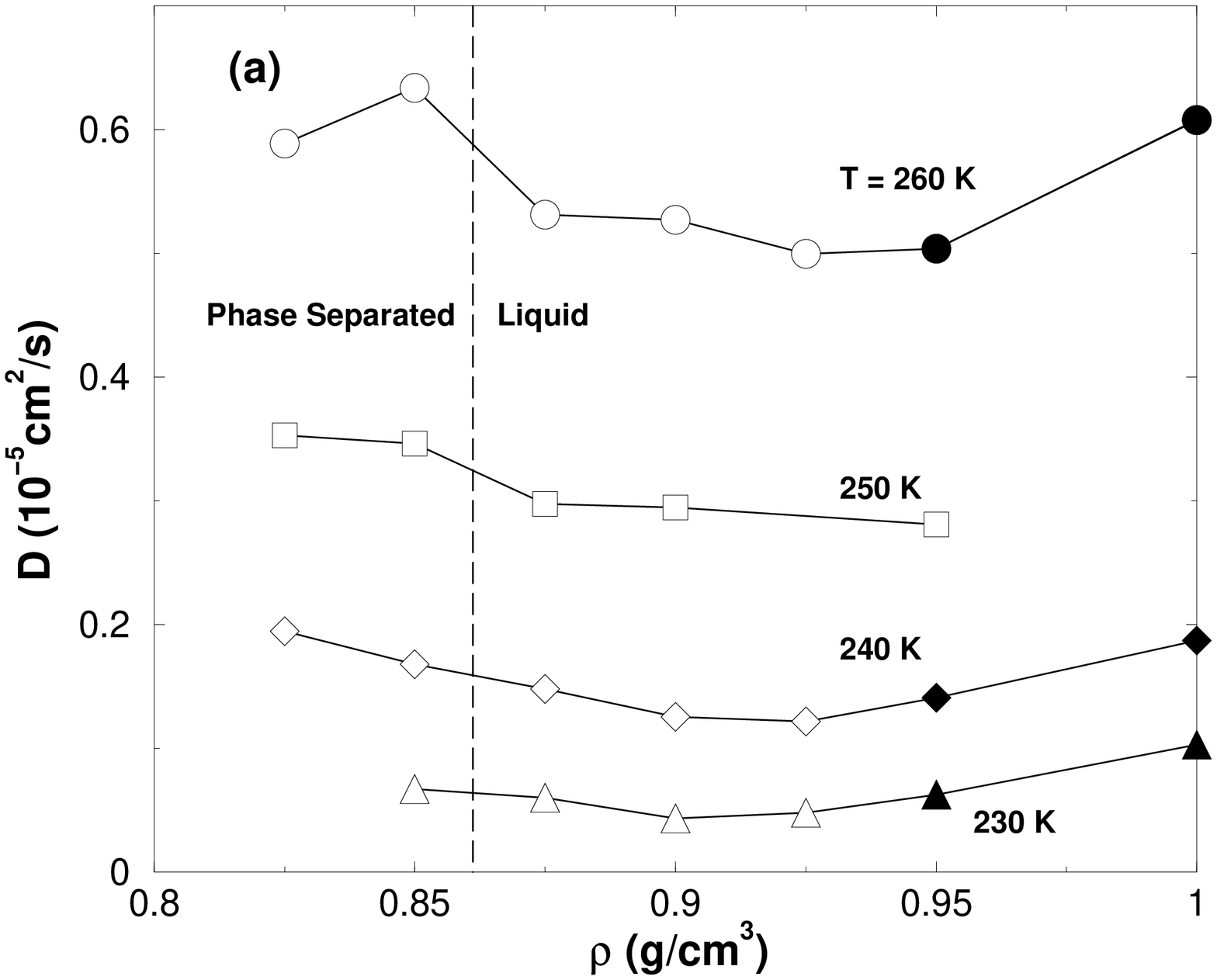}
\includegraphics[width=8.6cm,angle=-90]{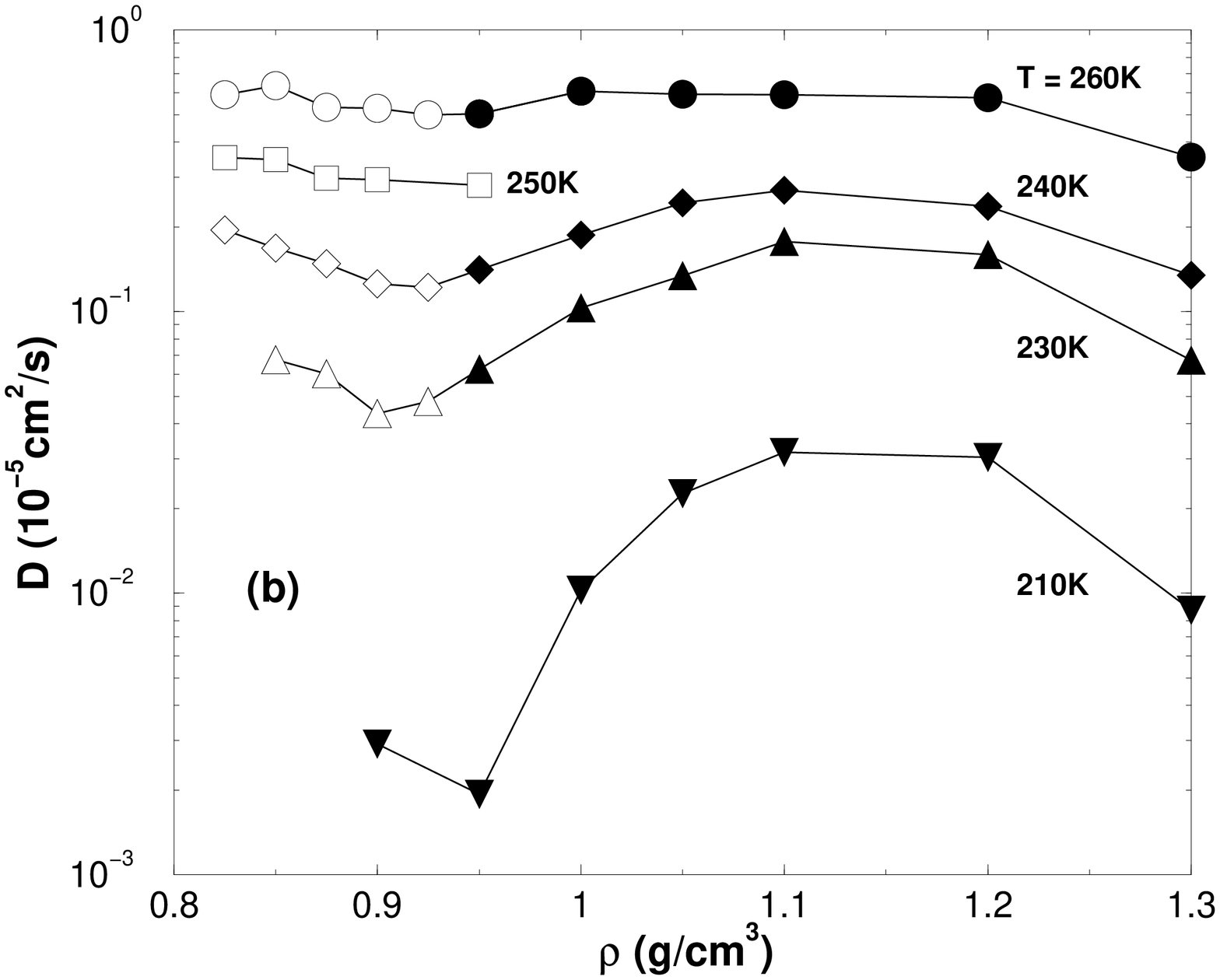}
\caption{(a) Dependence of the diffusion constant $D$ on $\rho$ along
  isotherms (for $\rho \le 1.0$~g/cm$^3$).  Open symbols are the new
  simulations we report, and filled symbols are from ref.~\cite{sta99e}.
  The dotted line separates liquid state points from phase separated
  state points, but is {\it not} an indication of the exact $\rho_{\rm
    sp}(T)$, which varies slightly with $T$. (b) Full $\rho$ dependence
  of $D$, also showing the maxima.}
\label{f4}
\end{figure}

Having established the limit of stability at negative pressures, we next
analyze the dynamics of stretched water.  The effect of extreme
conditions on the flow of the liquid we assess by calculating the
diffusion constant $D$, defined by the asymptotic value, obtained
by linear regression,  of the slope of
the mean square displacement versus time. We show $D$ along
isotherms in Fig.~\ref{f4}.  For $T\le260$~K, $D$ has a minimum value at
$\rho\approx 0.9$~g/cm$^3$, which becomes more pronounced at lower $T$
(Fig.~\ref{f4}b).  This behavior can be understood considering the
structural changes that occur with decreasing density. At low $T$, the
decreased density enhances the local tetrahedral ordering, which leads
to a decrease in $D$.  Further decreases in density reduces the
stability of the tetrahedral structure and causes an increase of $D$.

\begin{figure}
\includegraphics[width=8.6cm,angle=-90]{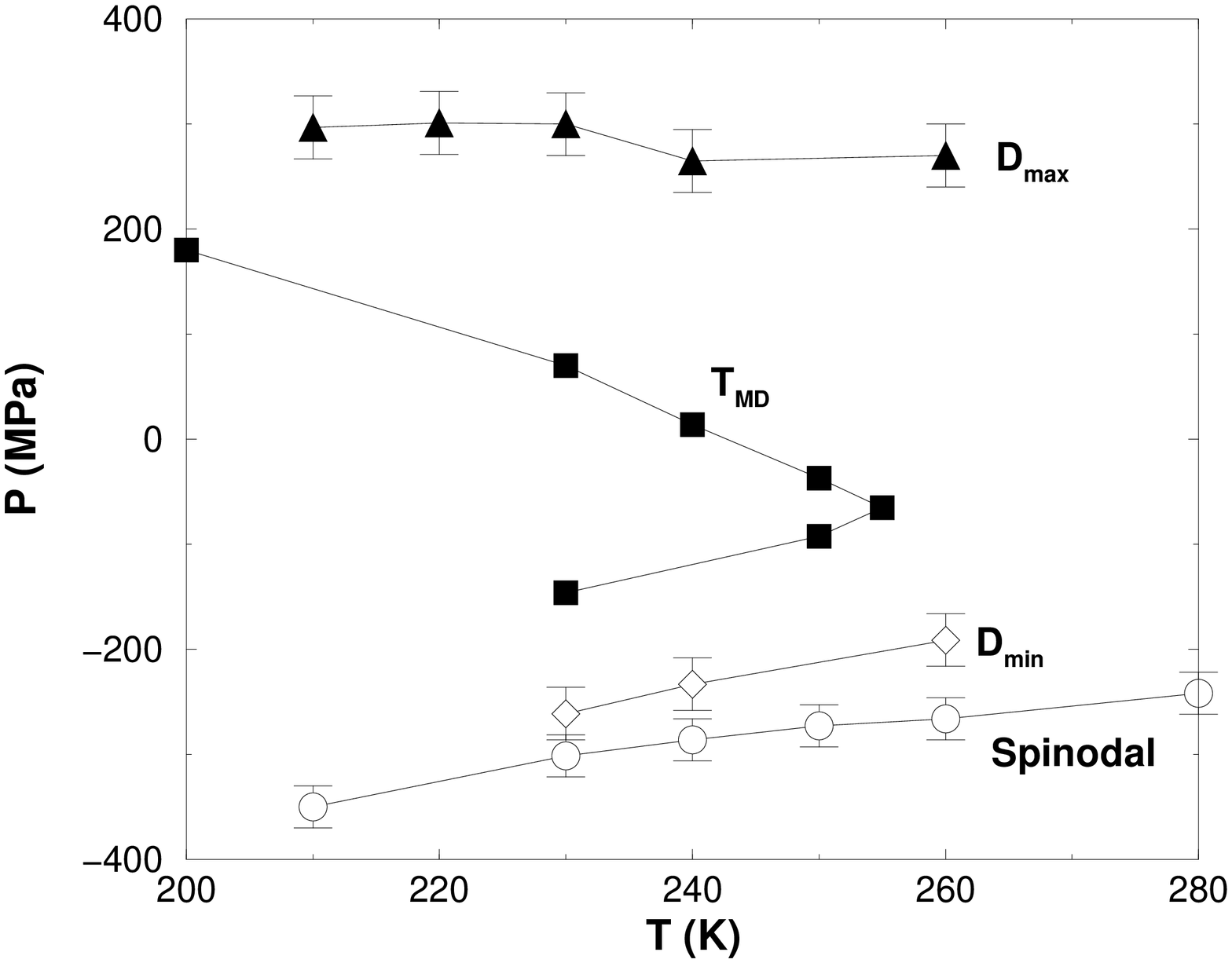}
\caption{Relation of the loci of maxima and minima of $D$ with 
$T_{MD}(P)$ and the spinodal.  Open symbols are from the present work, and
filled symbols are from ref.~\cite{sta99e}. There is no maximum and
minimum of $D$ for temperatures  $T\geq 280\;K$  Fig.~\ref{f4}} 
and ref.~\cite{sta99e} 

\label{loci}
\end{figure}

The location of the minimum is near the ice Ih density $\approx
0.915$~g/cm$^3$, which is the density where the perfect tetrahedral
order occurs.  The behavior of the minimum of $D$, $D_{\rm min}(T)$,
complements the known behavior of $D_{\rm max}(T)$ for the same model
\cite{sta99e,scala,err00}, where a maximum occurs due to breaking
hydrogen bonds at high pressure; the density of the $D_{\rm min}(T)$
increases slightly with increasing $T$, while the density of $D_{\rm
max}(T)$ decreases with increasing $T$ \cite{scala}.  This is expected,
since the range of densities where anomalous behavior occurs expands
with decreasing $T$.  We show the loci of $D_{\rm min}(T)$,  $D_{\rm
max}(T)$, along with the spinodal and the temperatures  of maxima density,
$T_{MD}(P)$ in 
Fig.~\ref{loci}.

\section{Conclusions}

Water exhibits a very complex structure and its properties and anomalies
are strongly influenced by variations of pressure.  For high densities
($\rho>\rho_{\mbox{\scriptsize max}}$), water behaves as a normal liquid
and the decrease of $D$ with increasing pressure is governed by steric
effects. For $\rho_{\mbox{\scriptsize min}} < \rho <
\rho_{\mbox{\scriptsize max}}$, as the pressure is decreased, the
presence of defects and interstitial water decrease, the tetrahedral
structure dominates, with stronger hydrogen bonds. This process reaches
its maximum at $\rho=\rho_{\mbox{\scriptsize min}} \approx
\rho_{\mbox{\scriptsize ice}}$. Further stretching destabilizes the
hydrogen bond network, leading to an increase in mobility. The locus of
$D_{\rm min}$ roughly tracks the spinodal, not surprising since the same
breakdown of tetrahedral order that gives rise to $D_{\rm min}$ also
facilitates cavitation.

\acknowledgments

We thank the National Science Foundation (NSF), the Conselho Nacional de
Desenvolvimento Cientifico e Technologico (CNPq), the Fundacao de Amparo
a Pesquisa do Rio Grande do Sul (Fapergs) for financial support.  FWS
thanks the National Research Council for financial support.

\begin{table}[tbp]
\begin{center} \caption{Temperature, density, diffusion constant, 
potential energy, pressure, equilibration $t_{eq}$ and production
$t_{pr}$times. The uncertainty in $D$ is roughly $\pm 0.008$.}
\label{jobs}
\begin{tabular}{|c|c|c|c|c|c|c|} \hline 
T (K) & $\rho$~(g/cm$^3$) & U (kJ/mol)& P (MPa) & $D$
($10^{-5}$~cm$^2$/s) & $t_{\mbox {\scriptsize eq}}$ (ns) & $t_{\mbox
{\scriptsize pr}}$ (ns) \\ \hline  
280 & 0.825 & -46.80 & -217 $\pm$ 18 & 1.237 & 1.5 & 1.5 \\
& 0.850 & -47.01 & -239 $\pm$ 19 & 1.359 & 1.5 & 1.5 \\
& 0.875 & -47.32 & -230 $\pm$ 17 & 1.281 & 2 & 2\\
& 0.900 & -47.75 & -204 $\pm$ 24 & 1.261 & 1.5 & 1.5\\
& 0.925 & -47.90 & -172 $\pm$ 29 & 1.234 & 1.5 & 1.5\\
& 0.950 & -48.12 & -133 $\pm$ 17 & 1.068 & 1.5 & 1.5\\
\hline 
260 & 0.825 & -48.27 & -221 $\pm$ 20 & 0.589 & 1.5 & 1.5\\ 
& 0.850 & -48.53 & -261 $\pm$ 17 & 0.634 & 1.5 & 1.5\\ 
& 0.875 & -48.88 & -257 $\pm$ 19 & 0.531 & 2 & 4\\
& 0.900 & -49.23 & -231 $\pm$ 20 & 0.527 & 2 & 4\\ 
& 0.925 & -49.48 & -191 $\pm$ 21 & 0.500 & 2 & 4\\
\hline 250 & 0.825 & -49.12 & -227 $\pm$ 11 & 0.353 & 2 & 2\\ 
& 0.850 & -49.28 & -273 $\pm$ 15 & 0.346 & 2 & 3\\ 
& 0.875 & -49.76 & -271 $\pm$ 18 & 0.298 & 2 & 3\\ 
& 0.900 & -50.03 & -244 $\pm$ 17 & 0.295 & 2 & 3\\
& 0.950 & -50.40 & -148 $\pm$ 18 & 0.281 & 1 & 1.5\\ 
\hline 240 & 0.825 & -49.99 & -230 $\pm$ 19 & 0.195 & 5 & 5\\ 
& 0.850 & -50.17 & -263 $\pm$ 27 & 0.168 & 5 & 5\\ 
& 0.875 & -50.51 & -282 $\pm$ 18 & 0.148 & 5 & 5\\ 
& 0.900 & -50.98 & -258 $\pm$ 17 & 0.126 & 5 & 5\\ 
& 0.925 & -51.30 & -213 $\pm$ 19 & 0.122 & 5 & 5\\
\hline 230 & 0.850 & -51.06 & -278 $\pm$ 16 & 0.0674& 5 & 5\\
& 0.875 & -51.28 & -300 $\pm$ 18 & 0.060 & 5 & 5\\
& 0.900 & -51.80 & -272 $\pm$ 16 & 0.044 & 5 & 5\\
& 0.925 & -52.01 & -212 $\pm$ 17 & 0.048 & 10 & 10\\
\hline 210 & 0.850 & -52.73 & -314 $\pm$ 15 & - & 10 & 10\\ 
& 0.875 & -52.92 &-348 $\pm$ 21 & - & 10 & 50\\ 
& 0.925 & -53.54 &-211 $\pm$ 24 & - & 10 & 10\\
\hline 
\end{tabular} 
\end{center}
\end{table}

\newpage

\begin{thebibliography}{99}

\bibitem{debe96}
P. G. Debenedetti, {\it Metastable Liquids\/} (Princeton University
Press, Princeton NJ, 1996).

\bibitem{ange82} 
O. Mishima and H. E. Stanley, {\it Nature\/} {\bf 396}, 329 (1998).

\bibitem{four97}
J. T. Fourkas, D. Kivelson, U. Mohanty, and K. A. Nelson, eds., {\it
Supercooled Liquids: Advances and Novel Applications\/} (ACS Books,
Washington DC, 1997).

\bibitem{sta97a}
H. E. Stanley, L. Cruz, S. T. Harrington, P. H. Poole, S. Sastry,
F. Sciortino, F. W. Starr, and R. Zhang, {\it Physica A\/} {\bf 236}, 19
(1997).


\bibitem{lan82}
E. W. Lang and H.-D. L\"udemann, {\it Angewandte Chemie, International
Edition in English\/} {\bf 21}, 315 (1982).

\bibitem{dou98}
R. C. Dougherty and L. N. Howard, {\it J. Chem. Phys.} {\bf 109}, 7379
(1998).

\bibitem{geig79}
A. Geiger, F. H. Stillinger, and A. Rahman, 
{\it J. Chem. Phys.} {\bf 70}, 4185 (1979).

\bibitem{Stanley80}
H. E. Stanley and J. Teixeira, 
{\it J. Chem. Phys.} {\bf 73}, 3404 (1980).

\bibitem{spe82b}
R. J. Speedy, 
{\it J. Chem. Phys.} {\bf 86}, 982 (1982); {\it Ibid} {\bf 86}, 3002 (1992).

\bibitem{spe87}
R. J. Speedy, 
{\it J. Chem. Phys.} {\bf 91}, 3354 (1987).


\bibitem{pses} P.~H. Poole, F. Sciortino, U. Essmann, and H.~E. Stanley,
Nature {\bf 360}, 324 (1992); Phys. Rev. E {\bf 48}, 3799 (1993);
F. Sciortino, P.H. Poole, U. Essmann, and H.E. Stanley, Ibid {\bf 55},
727 (1997); S. Harrington, R. Zhang, P.H. Poole, F. Sciortino, and
H.E. Stanley, Phys. Rev. Lett. {\bf 78}, 2409 (1997).

\bibitem{mishima94}
O. Mishima, J. Chem. Phys. {\bf 100},  5910  (1994).

\bibitem{critical-point}
C.~J. Roberts, A.~Z. Panagiotopoulos, and P.~G. Debenedetti,
Phys. Rev. Lett.  {\bf 97}, 4386 (1996); C.~J. Roberts and
P.~G. Debenedetti, J. Chem. Phys. {\bf 105}, 658 (1996).

\bibitem{ms98}
M.-C. Bellissent-Funel, Europhys. Lett. {\bf 42}, 161 (1998); O. Mishima
and H.~E. Stanley, Nature {\bf 392}, 192 (1998).

\bibitem{tana96}
H. Tanaka, 
{\it J. Chem. Phys.} {\bf 105}, 5099 (1996).


\bibitem{t35}
A. Scala, F. W. Starr, E. La Nave, H. E. Stanley and
F. Sciortino, {\it  Phys. Rev. E} {\bf 62},  8016 (2000).


\bibitem{sas96}
S. Sastry, P. G. Debenedetti, F. Sciortino, and H. E. Stanley,
{\it Phys. Rev. E\/} {\bf 53}, 6144 (1996).

\bibitem{sas98}
L. P. N. Rebelo, P. G. Debenedetti, and S. Sastry, 
{\it J. Chem. Phys.} {\bf 109}, 626 (1998).

\bibitem{net98a}
P. A. Netz and Th. Dorfm\"uller, 
{\it J. Phys. Chem. B\/} {\bf 102}, 4875 (1998).

\bibitem{koga98}
K. Koga, X. C. Zeng, and H. Tanaka, 
{\it Chem. Phys. Lett.} {\bf 285}, 278 (1998).

\bibitem{poc95}
W. T. Pockman, J. S. Sperry, and J. W. O'Leary, 
{\it Nature\/} {\bf 378}, 715 (1995).

\bibitem{jon76}
J. Jonas, T. DeFries, and D. J. Wilbur, 
{\it J. Chem. Phys.}  {\bf 65}, 582 (1976).

\bibitem{pri87}
F. X. Prielmeier, E. W. Lang, R. J. Speedy, and H.-D. L\"udemann,
{\it Phys. Rev. Lett.}  {\bf 59}, 1128 (1987);
{\it Ber. Bunsenges. Phys. Chem.} {\bf 92}, 1111 (1988).

\bibitem{ram87}
M. Rami Reddy and M. Berkovitz, 
{\it J. Chem. Phys.} {\bf 87}, 6682 (1987).

\bibitem{sci91}
F. Sciortino, A. Geiger, and H. E. Stanley, 
{\it Nature\/} {\bf 354}, 218 (1991); Ibid. 
{\it J. Chem. Phys.} {\bf 96}, 3857 (1992).

\bibitem{bae94}
L. A. Baez and P. Clancy, 
{\it J. Chem. Phys.} {\bf 101}, 9837 (1994).

\bibitem{har97}
S. Harrington, P. H. Poole, F. Sciortino, and H. E. Stanley, 
{\it J. Chem. Phys.}  {\bf 107}, 7443 (1997).




\bibitem{francesco} P. Gallo, F. Sciortino, P. Tartaglia, and
S.-H. Chen, Phys. Rev. Lett. {\bf 76}, 2730 (1996); F. Sciortino,
P. Gallo, P. Tartaglia, S.-H. Chen, Phys. Rev. E {\bf 54}, 6331 (1996);
S.-H. Chen, P. Gallo, F. Sciortino, and P. Tartaglia, {\it Ibid} {\bf
56}, 4231 (1997); F. Sciortino, L. Fabbian, S.-H. Chen, and
P. Tartaglia, {\it Ibid}, 5397 (1997).

\bibitem{sta99e}
F.W. Starr, F. Sciortino, and H.E. Stanley, 
{\it Phys. Rev. E\/} {\bf 60}, 6757 (1999); F.W. Starr, S.T. Harrington,
F.~Sciortino, and H.E. Stanley, {\em Phys. Rev. Lett.}, {\bf 82}, 3629,
(1999).

\bibitem{scala}
A. Scala, F. W. Starr, E. La Nave, F. Sciortino and H. E. Stanley,
Nature  {\bf 406}, 166 (2000). 





\bibitem{hen80}
S. J. Henderson and R. J. Speedy, 
{\it J. Phys. E: Scientific Instrumentation\/} {\bf 13}, 778 (1980).

\bibitem{GDWA90} 
J. L. Green, D. J. Durben, G. H. Wolf, and C. A. Angell, {\it Science\/}
{\bf 249}, R649 (1990).

\bibitem{vai93}
I. I. Vaisman, L. Perera, and M. L. Berkovitz, 
{\it J. Chem. Phys.} {\bf 98}, 9859 (1993).







\bibitem{ge94}P. Mousbach, J. Schnitker and  A. Geiger, {\it Z. Angew Math.
Mech} {\bf 74}, T608 (1994).




\bibitem{err00}
J. R. Errington and P. G. Debenedetti,
{\it Nature} {\bf 409}, 318  (2001).

\bibitem{g(r)note}
Note that while $S(q)$ clearly indicates phase separation in the system,
the static pair correlation function $g(r)$ (the Fourier transform of
$S(q)$) in the unstable region of the phase diagram shows no significant
difference from $g(r)$ in the metastable region.



\end{thebibliography}
\end{document}